\newcommand{\plot}[2]{\centering \leavevmode
    \epsfxsize=#2\textwidth \epsfbox{#1}}

\def\hii{\hbox{H~{\scriptsize II}}}
\def\arcsec{\hbox{$^{\prime\prime}$}}
\def\farcs{\hbox{$.\!\!^{\prime\prime}$}}

\def\ha{\hbox{H$\alpha$}}

\def\hii{\hbox{H~{\scriptsize II}}}

\def\nii{\hbox{[N~{\scriptsize II}]}}
\def\sii{\hbox{[S~{\scriptsize II}]}}
\def\oi{\hbox{[O~{\scriptsize I}]}}

\documentstyle[epsf]{cupconf}

\title[Central UV Spikes in two Galactic Spheroids]
{Central UV Spikes in two Galactic Spheroids\thanks{Based on observations
with the NASA/ESA Hubble Space Telescope, obtained at the Space Telescope
Science Institute, which is operated by AURA, Inc., under NASA Contract NAS
5-26555.}}

\author[M. Cappellari {\em et al.}]{Michele~Cappellari$^2$,
    Francesco~Bertola$^2$, David~Burstein$^3$,
    Lucio~M.~Buson$^4$, Laura~Greggio$^{5,6}$,
    Alvio~Renzini$^{7,8}$}

\affiliation{$^2$Dipartimento di Astronomia,
    	Universit\`a di Padova, Padova, Italy\\[\affilskip]
	$^3$Department of Physics \& Astronomy,
    	Arizona State University, USA\\[\affilskip]
	$^4$Osservatorio di Capodimonte, Napoli, Italy\\[\affilskip]
	$^5$Osservatorio di Bologna, Bologna, Italy\\[\affilskip]
	$^6$Universitaets Sternwarte, Muenchen, Germany\\[\affilskip]
	$^7$Dipartimento di Astronomia,
    	Universit\`a di Bologna, Bologna, Italy\\[\affilskip]
	$^8$European Southern Observatory,
    	Garching bei M\"unchen, Germany}

\begin{document}

\maketitle

\begin{abstract}
FOS spectra and FOC photometry of two centrally located, UV-bright spikes
in the elliptical galaxy NGC~4552 and the bulge-dominated early spiral
NGC~2681, are presented. These spectra reveal that such point-like UV
sources detected by means of HST within a relatively large fraction
($\sim15$\%) of bulges can be related to radically different phenomena.
While the UV unresolved emission in NGC~4552 represents a transient event
likely induced by an accretion event onto a supermassive black hole, the
spike seen at the center of NGC~2681 is not variable and it is stellar in
nature.
\end{abstract}

\firstsection

\section{Introduction}

HST UV images of nearby galaxies presented by Maoz et al. (1996) and Barth
et al. (1998), as well as analogous space-borne optical images of
early-type galaxies discussed by Lauer et al. (1995) and Carollo et al.
(1997) have shown that about 15\% of imaged galaxies show evidence of
unresolved central spikes.

In the following we discuss two `prototype' galactic spheroids---NGC~2681
and NGC~4552 we properly monitored with HST---which host UV-bright,
unresolved spikes at their center. While the early-spiral (Sa) galaxy
NGC~2681 shows a {\em nonvariable} unresolved cusp, the UV spike which
became visible at the center of the Virgo Elliptical NGC~4552 is a UV flare
caught in mid-action, presumably related to a transient accretion event
onto a central supermassive black hole (Renzini et al. 1995; Cappellari et
al. 1998).

Although radically different phenomenologies are involved, the appearance
of either nuclei---recently imaged in the UV (FOC/96 F342W) by means of the
refurbished HST---is quite similar. Nevertheless, basic pieces of
information can still be extracted from photometric profiles alone which
represent a potential diagnostics to disentangle the above scenarios. For
instance, the UV-bright unresolved spike observed at the center of NGC~2681
{\it does not} vary and matches a pure nuker law profile of the power law
type (Cappellari et al.\ 1999). On the contrary, in order to model the
flaring UV spike at the center of NGC~4552 one has to add to the observed
galaxy profile the contribution of an unresolved central point source,
whose intensity is allowed to vary (see Fig.~1).

\begin{figure}
\plot{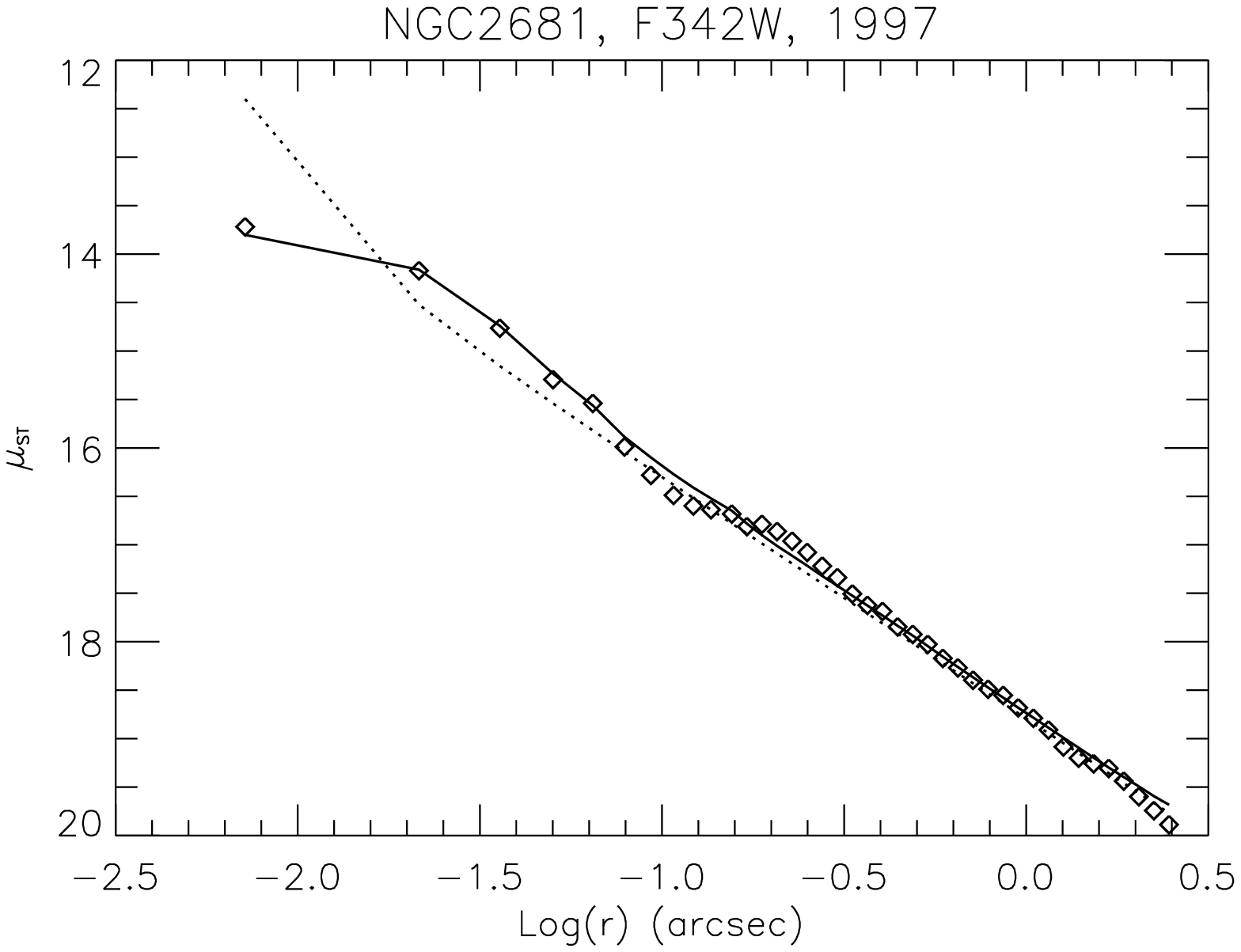}{.45}\plot{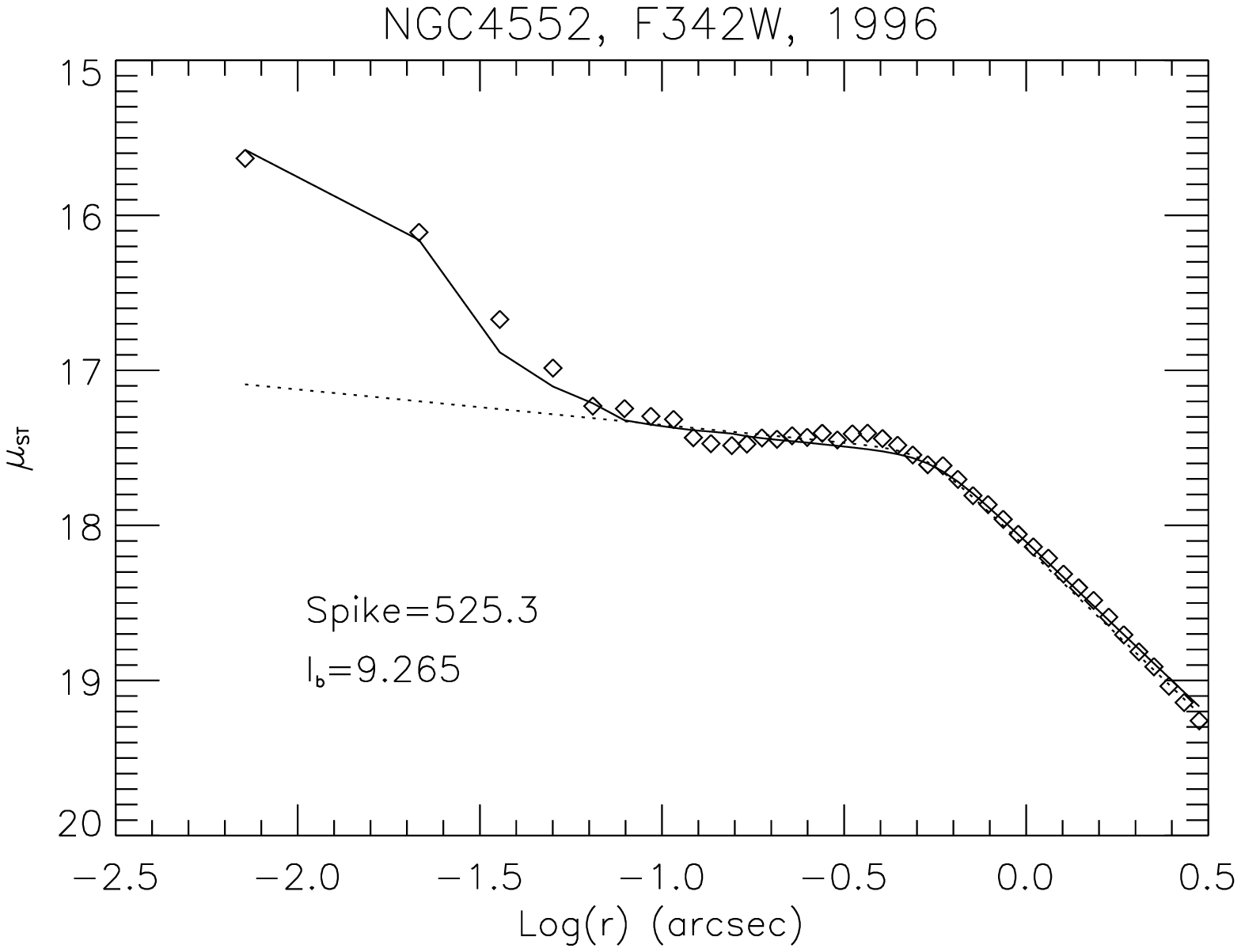}{.45}
\caption{The inner surface brightness profiles (arbitrary magnitudes) of
	NGC~2681 (Left panel) and NGC~4552 (Right panel) vs. $\log r$ in the
	FOC/96 F342W waveband. Diamonds represent the observed profiles, the
	dashed lines represent models of the true galaxy profiles and finally
	the solid line shows the above models after convolving with the proper
	PSF. NGC~2681 is a typical power-law bulge, while NGC~4552 is a classic
	giant elliptical with core, but one need to add a central point-like
	source to the Nuker-law in order to reproduce the observed profile.}
\end{figure}

\section{Observations and Reductions}

FOC UV observations of NGC 4552 obtained in 1991, 1993 and 1996 are
described in detail by Cappellari et al. (1998). These data include a
single FOC/96 F342W frame obtained on July 19, 1991 and subsequent images
obtained on November 27-28, 1993 in four consecutive UV passbands (FOC/96
F175W, F220W, F275W, F342W). We observed NGC~4552 for a third time on May
24, 1996 with COSTAR-Corrected HST making use of a comparable set of UV
filters as in 1993 (FOC/96 F175W, F275W and F342W). Initial FOC images of
NGC 2681 were obtained by our group on November 4-5, 1993 in the FOC F175W,
F220W, F275W, and F342W filters, pre-COSTAR. As with NGC 4552, we also
obtained a set of post-COSTAR UV images on February 1, 1997 of NGC 2681
with the same FOC filter set (apart from F220W) as used in 1993. All FOC
images have been re-calibrated in a self-consistent manner, including all
required correction factors for PSF and sensitivity differences
(zoom/non-zoomed modes and COSTAR) as well as nonlinearity effects. In
addition to the FOC images obtained in 1996, we were also able to obtain
FOS spectra of both galaxies. The FOS peak-up procedure was used to locate
the 0\farcs2 square aperture on the nucleus of each galaxy (as confirmed
via the multiple peak-up output). FOS gratings G270H, G650L and G780H were
used for each galaxy. The nuclear spectra of NGC~4552 and NGC~2681 were
obtained on May 24, 1996 and on February 2, 1997, respectively.

\begin{figure}
\plot{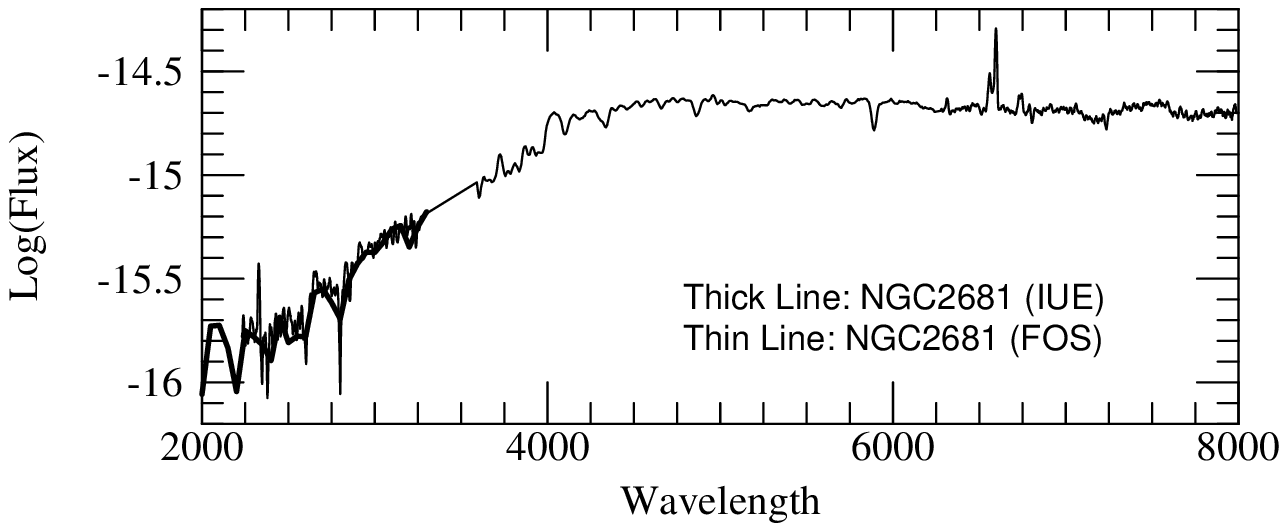}{.83}

\plot{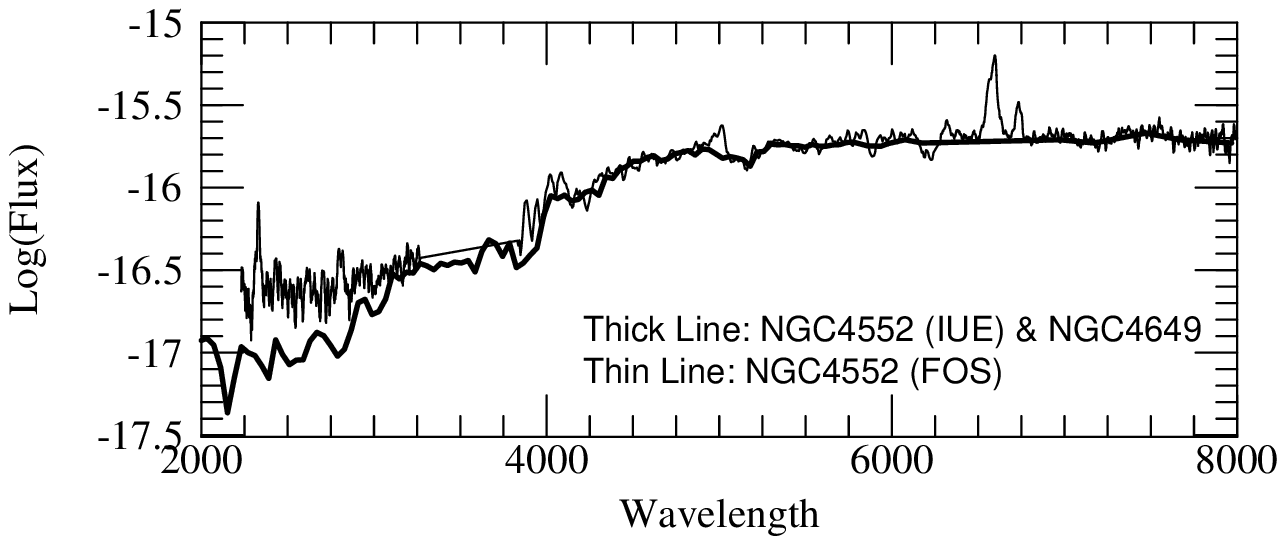}{.83}
\caption{{\em Upper panel:} The overall 1997 FOS spectrum of NGC~2681
	within the $0\farcs2\times0\farcs2$ aperture centered on the spike
	(thin line), is superimposed to the IUE spectrum of the same galaxy
	within a 10\arcsec$\times$20\arcsec\ aperture (Burstein et al.\ 1988),
	properly normalized to the visual region. {\em Lower panel:} The 1996
	FOS spectrum of NGC~4552 centered on the spike (thin  line), is
	superimposed to a scaled combination of the IUE spectrum of the same
	galaxy matched to ground-based optical spectrum of NGC~4649, a giant
	elliptical whose SED is virtually the same as that of NGC~4552 (thick
	line). The spectra have been normalized to the visual region. The
	NGC~4552 spectrum appears quite different owing to a continuum UV
	excess shortward of $\lambda\sim$3000~\AA. This UV excess is absent in
	NGC~2681.}
\end{figure}

\begin{figure}
\plot{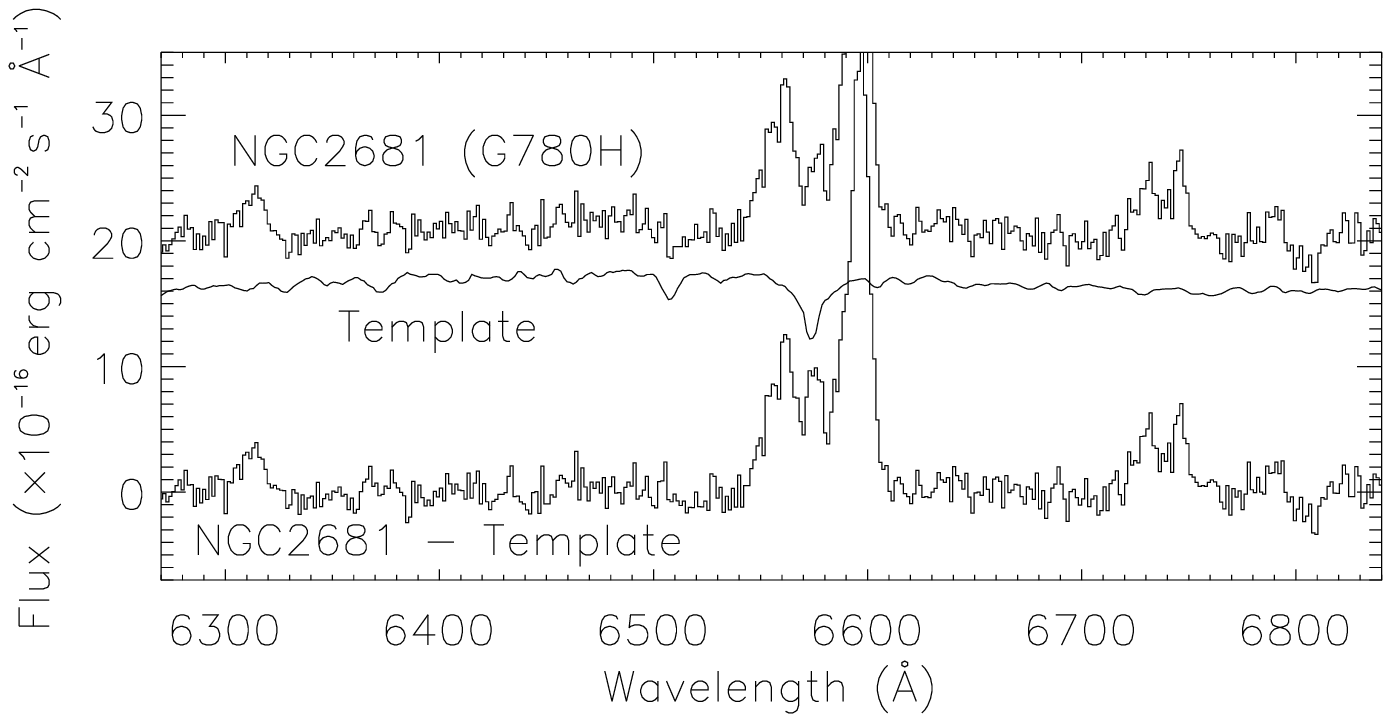}{.5}\plot{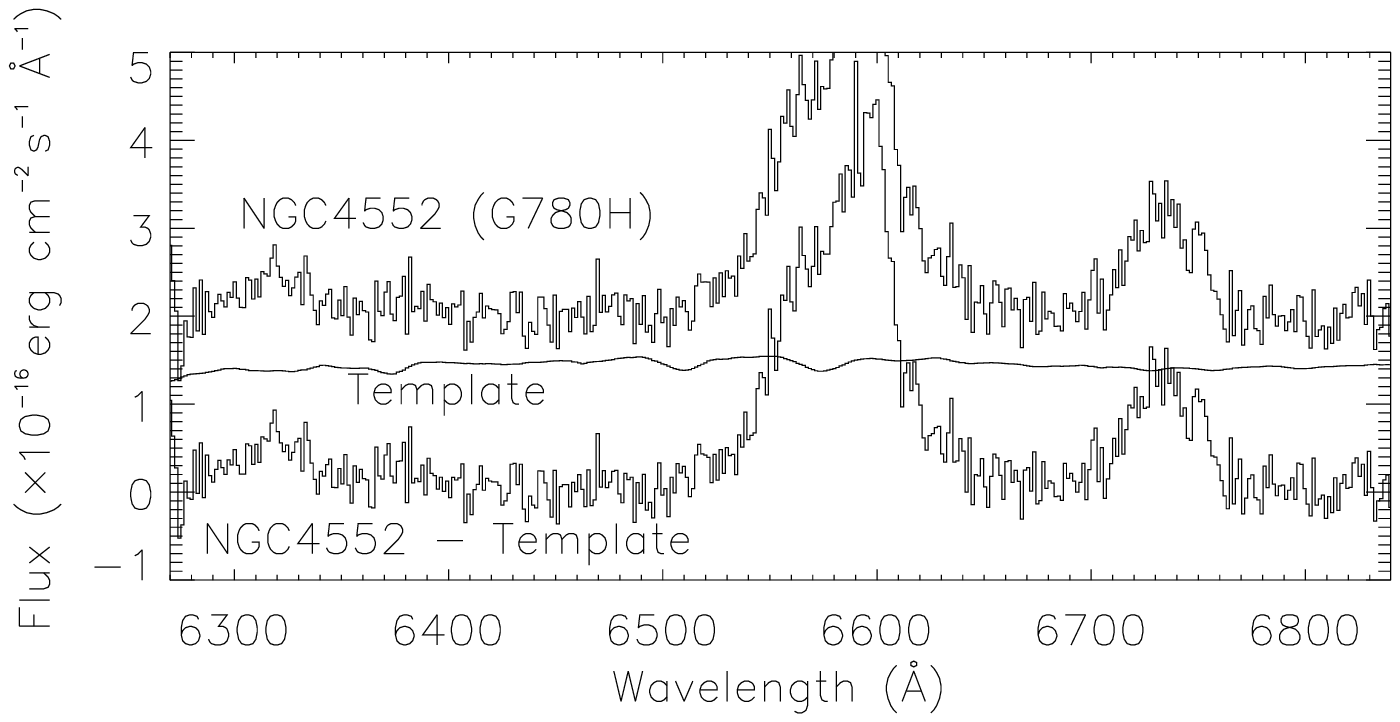}{.5}
\caption{The FOS G780H spectrum of NGC~2681 (left panel) and NGC~4552
	(right panel), showing the region of the \oi, \nii, \ha, and \sii\
	emission lines. In both panels a starlight template (obtained as in Ho
	et al. 1997) has been subtracted from the upper plot spectrum to obtain
	the continuum-free spectrum of the lower plot.}
\end{figure}

\section{Results}

The ultraviolet-bright source in NGC~4552 was first detected in 1991, it
increased in luminosity by a factor of $\sim 4.5$ by 1993, and then
declined a factor of $\sim 2.0$ by 1996, On the contrary 1993 and 1997 UV
FOC observations of NGC~2681 are consistent with no variation at all.

The overall nuclear FOS spectra of NGC~2681 and NGC~4552, together with the
IUE and optical underlying spectra normalized to the visual region, are
shown in Fig.~2. As is evident, in the case of NGC~2681 the match of the
two spectra is quite striking, thus implying that the UV continuum flux of
NGC~2681 is simply dominated by its stellar population, essentially
identical in its innermost regions and in the whole
10\arcsec$\times$20\arcsec\ IUE aperture. On the other hand, FOS
spectroscopy of NGC~4552 reveals a strong UV continuum {\em over} the
spectrum of the underlying galaxy, along with several emission lines in
both the UV and the optical ranges. The SED of the spike alone---obtained
by subtracting the V-mag normalized IUE spectrum of the galaxy from the FOS
spectrum---indicates a temperature of $T\sim15000$ K for the spike in 1996,
if a thermal origin for the UV flux is assumed.

\begin{figure}
\plot{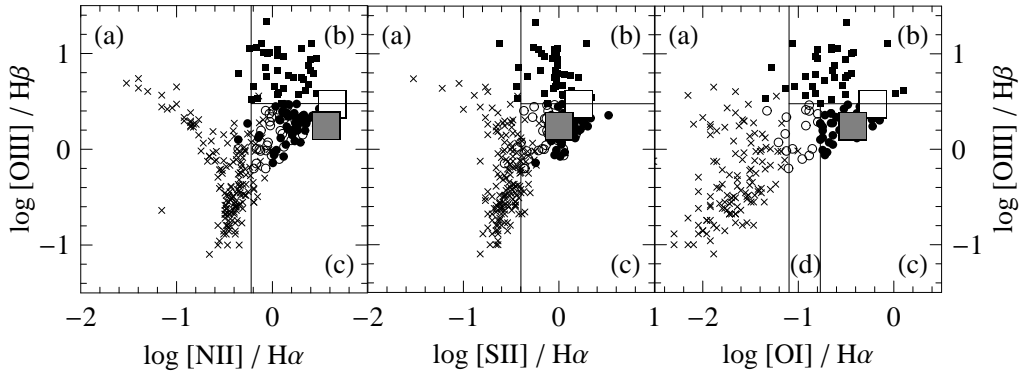}{1}
\caption{The location of the NGC~2681 (large gray square) and NGC~4552
	(large open square) nucleus (as derived from the narrow line emission
	components measured on the FOS spectra) on the diagnostic diagrams used
	by Ho et al.\ (1997). The corresponding errors are of the size of the
	smaller simbols. The other symbols represent the nuclei included in the
	Ho et al. sample (crosses = \hii\ nuclei, filled squares = Seyfert
	nuclei, filled circles = LINERs, open circles = transition objects).
	The vertical and horizontal lines delineates the boundary adopted by Ho
	et al.\ between (a) \hii\ nuclei, (b) Seyfert galaxies, (c) LINERS and
	(d) Transition objects.}
\end{figure}

The FOS G780H spectra of NGC~2681 and NGC~4552 including the \oi, \nii,
\ha, and \sii\ emission lines are presented in Fig.~3. In NGC~2681 all
lines are well fitted by a single gaussian component with FWHM of $\sim470$
km s$^{-1}$. In the case of NGC~4552 however both permitted {\it and}
forbidden lines are best modelled with a combination of broad and narrow
components, with FWHM of $\sim 3000$ km s$^{-1}$ and $\sim 700$ km
s$^{-1}$, respectively. The 1996 broad H$\alpha$ luminosity of this
mini-AGN is $\sim 5.6\times 10^{37}$ erg s$^{-1}$, about a factor of two
less than that of the nucleus of NGC~4395, heretofore considered to be the
faintest known AGN (Filippenko et al. 1993).

The FOS spectroscopy indicates also a significant similarity between the
two nuclei, namely their emission line ratios and related gas diagnostics
and UV-source classification. A comparison of the emission line ratios of
the narrow components for both the NGC~4552 and NGC~2681 spikes with the
distribution of Seyfert galaxies, LINERS and \hii\ regions in the
diagnostic emission line diagrams of Ho et al.\ (1997) is given in Fig.~4.
As is evident, the line ratios definitively place both spikes among extreme
AGNs. The ratios for NGC~4552 fall just on the borderline between Seyferts
and LINERs, while those measured for NGC~2681 indicate that this nucleus
can be classified as a LINER.

\end{document}